\documentclass[sigconf,screen]{acmart}

\setcopyright{cc}
\setcctype{by}
\acmDOI{10.1145/3832783.3834630}
\acmYear{2026}
\copyrightyear{2026}
\acmISBN{979-8-4007-2882-2/2026/10}
\acmConference[ASE '26]{Proceedings of the 41st IEEE/ACM International Conference on Automated Software Engineering}{October 12--16, 2026}{Munich, Germany}
\acmBooktitle{Proceedings of the 41st IEEE/ACM International Conference on Automated Software Engineering (ASE '26), October 12--16, 2026, Munich, Germany}
\acmSubmissionID{ase26tool-p86-p}
\received{2026-05-12}
\received[accepted]{2026-06-19}
\makeatletter
\AtEndPreamble{%
  \global\@ACM@balancefalse
  \RequirePackage{pbalance}
}
\makeatother

\AtBeginDocument{%
  
}
\usepackage{flushend}
\usepackage[inline]{enumitem}
\usepackage{booktabs}

\usepackage{amsmath,amssymb}
\usepackage{algorithm}
\usepackage{array}
\usepackage{algpseudocode}
\usepackage{graphicx}
\usepackage{textcomp}
\usepackage[dvipsnames]{xcolor}
\usepackage{minted}
\usepackage{tabularx}
\usepackage{listings}
\usepackage{multirow}
\usepackage[T1]{fontenc}
\usepackage{mathtools}
\usepackage{lipsum}
\usepackage{subcaption}
\usepackage{tcolorbox}
\usepackage{anyfontsize}
\usepackage{mdframed}
\usepackage{tikz}
\usepackage{makecell}
\usepackage[normalem]{ulem}
\usepackage{hyperref}
\usetikzlibrary{shapes.geometric, arrows.meta, positioning, fit, calc, backgrounds}

\definecolor{dkgreen}{rgb}{0,0.6,0}
\definecolor{codegray}{rgb}{0.5,0.5,0.5}
\definecolor{mauve}{rgb}{0.58,0,0.82}
\definecolor{mygray}{rgb}{0.9,0.9,0.9}

\DeclareUnicodeCharacter{2212}{-}

\begin{document}

\title{MergeSE: Post-Hoc Model Merging for Software Engineering Tasks Without Retraining}

\title{MergeSE: Post-Hoc Model Merging for Software Engineering Tasks without Retraining}

\author{Palash R. Roy}
\correspondingauthor
\orcid{0000-0001-9470-4233}
\affiliation{%
  \institution{University of Saskatchewan}
  \department{Computer Science}
  \city{Saskatoon}
  \country{Canada}
}
\email{palash.roy@usask.ca}

\author{Banani Roy}
\orcid{0000-0003-1247-7781}
\affiliation{%
  \institution{University of Saskatchewan}
  \department{Computer Science}
  \city{Saskatoon}
  \country{Canada}
}
\email{banani.roy@usask.ca}

\author{Kevin A. Schneider}
\orcid{0000-0003-1113-1754}
\affiliation{%
  \institution{University of Saskatchewan}
  \department{Computer Science}
  \city{Saskatoon}
  \country{Canada}
}
\email{kevin.schneider@usask.ca}

\author{Chanchal K. Roy}
\orcid{0000-0003-0519-6164}
\affiliation{%
  \institution{University of Saskatchewan}
  \department{Computer Science}
  \city{Saskatoon}
  \country{Canada}
}
\email{chanchal.roy@usask.ca}

\begin{abstract}
Fine-tuned code models often behave as domain specialists and can degrade sharply under distribution shift: in our clone-detection setting, a model trained on same-language clones drops 71\% F1 on cross-language clones, while multi-task training falls to 0.151 F1 on unseen AI-generated clones. Our companion study shows that post-hoc model merging can address this fragmentation, achieving 93\% of multi-task performance without training data while generalizing 4$\times$ better to unseen clone types. However, no practical tool exists that lets SE researchers diagnose checkpoint compatibility, merge specialists, validate results on SE benchmarks, and export models for deployment. We present \textbf{MergeSE}, an open-source CLI and web tool for training-free model merging of HuggingFace encoder checkpoints. While motivated by OOD generalization in clone detection, MergeSE supports SE classification workflows more broadly through a built-in registry of nine task types, including vulnerability detection, defect prediction, and code-smell detection. MergeSE provides five operations: \textit{tasks}, \textit{inspect}, \textit{merge}, \textit{evaluate}, and \textit{export}. It supports five merging algorithms, including TIES, DARE-TIES, Wudi, PCB, and averaging; detects cross-task classification-head mismatches; produces seedable deterministic outputs; and includes bundled benchmark samples for smoke-test reproduction. A full merge of two 124M-parameter checkpoints completes in under 5 seconds on CPU. End-to-end validation confirms that MergeSE-produced checkpoints match reference implementations and recover cross-domain performance from domain-specific specialists. The tool is available online at \url{https://mergese.usask.ca}, and the development repository is at \url{https://github.com/srlabUsask/MergeSE}.

\end{abstract}


\maketitle

\section{Introduction}
\label{sec:introduction}
 
Software engineering researchers routinely fine-tune pre-trained code models such as CodeBERT~\cite{feng2020codebert}, GraphCodeBERT~\cite{guo2021graphcodebert}, and UniXcoder~\cite{guo2022unixcoder} on specific tasks, domains, and benchmarks. The result is a growing collection of specialist checkpoints, each effective in its own setting but unreliable outside it. A clone detector trained on BigCloneBench~\cite{svajlenko2014bigclonebench} may achieve 0.940~F1 on same-language clones yet collapse to 0.269~F1 on cross-language clones~\cite{roy2026unified}. Similar fragmentation is expected in other SE classification settings, such as vulnerability detection~\cite{du2024generalization}, defect prediction, and code-smell detection, where a model trained on one distribution may fail on another. This \textit{out-of-distribution (OOD) fragmentation} forces researchers and practitioners to maintain multiple models for closely related capabilities, with no practical way to combine them after fine-tuning.
 
The standard remedy, multi-task training on pooled data, requires simultaneous access to every training corpus, full retraining whenever a new domain appears, and careful balancing to avoid negative transfer. In our companion study~\cite{roy2026unified}, multi-task training on 1.38M clone pairs achieves 0.927 combined~F1 in-domain but collapses to 0.151~F1 on unseen GPT-generated clones~\cite{alam2023gptclonebench,roy2023unveiling}, worse than every individual specialist. This result suggests that joint training can overfit to the combined training distribution rather than improve robustness. Model merging~\cite{ilharco2022editing,yadav2023ties,yu2024language} offers a fundamentally different path. Instead of retraining, it combines trained checkpoints by operating directly on their parameters, requiring no training data, no gradient computation, and no GPU. Our companion study demonstrates that same-base TIES merging achieves 93\% of multi-task performance while generalizing 4$\times$ better to unseen clone types, and identifies task-vector compatibility through a shared pre-trained base as a key condition for successful merging.
 
These findings establish \textit{what} works. What is missing is a practical tool that lets SE researchers \textit{do} it. Existing model-merging implementations are often experiment-specific scripts with hard-coded paths, limited diagnostics, and no SE-oriented evaluation workflow. General-purpose tools such as mergekit~\cite{goddard2024arcee} are designed primarily around large language model merging workflows and do not provide SE-specific support for encoder-based code-model checkpoints, task registries, clone-pair evaluation, or benchmark validation. An SE researcher who wants to reuse two fine-tuned CodeBERT or UniXcoder checkpoints must currently answer several operational questions manually: Are the checkpoints architecturally compatible? Do their tokenizers match? Are their task vectors aligned enough to merge? Does the merged model preserve performance on SE benchmarks? Can the resulting checkpoint be exported for deployment?
 
We present \textbf{MergeSE}, an open-source CLI and web tool that operationalizes training-free model merging for software engineering. MergeSE provides five operations: \textit{tasks}, which lists supported SE task types; \textit{inspect}, which diagnoses compatibility using architecture metadata, tokenizer  checks, and task-vector geometry; \textit{merge}, which supports five merging algorithms including TIES~\cite{yadav2023ties}, DARE-TIES~\cite{yu2024language}, Wudi~\cite{cheng2025wudi}, PCB~\cite{du2024parameter}, and averaging without training data; \textit{evaluate}, which validates checkpoints on built-in or user-provided benchmarks using standard SE metrics; and \textit{export}, which packages merged models in HuggingFace,  ONNX, or TorchScript formats. While motivated by the OOD problem  in clone detection, MergeSE is task-general: it ships with a registry of nine SE classification tasks and supports any HuggingFace model, because the merging algorithms operate on model state dictionaries rather than task-specific data. A full merge of two 124M-parameter checkpoints completes in under 5~seconds on CPU.
 

\section{Tool Design}
\label{sec:design}
 
\begin{figure}[t]
\centering
\includegraphics[width=\columnwidth]{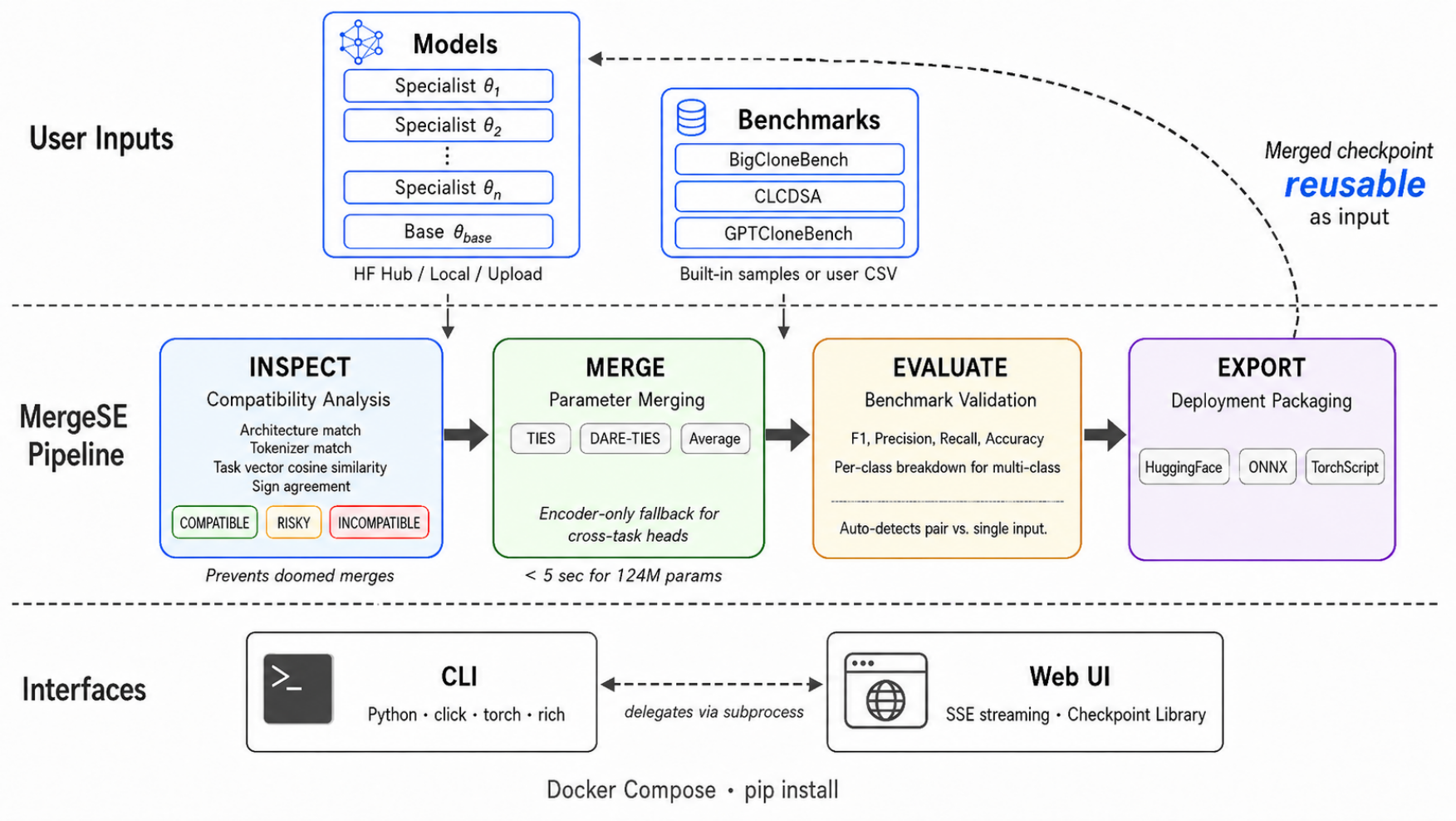}
\caption{MergeSE system architecture.}
\label{fig:architecture}
\end{figure}

Figure~\ref{fig:architecture} presents the overall architecture of 
MergeSE. The tool is organized around three design principles and 
five core operations, exposed through both a command-line interface 
and a web interface.
 
\noindent\textbf{Design Principles.}
MergeSE is guided by three design principles. First, \textit{no training and no GPU}: all merging operations are implemented as tensor arithmetic over existing checkpoints, making MergeSE usable on CPU-only machines and lowering the barrier for SE researchers who want to reuse fine-tuned models without access to large compute resources. Second, \textit{clone-motivated but SE-general}: MergeSE implements the practical recipe identified in our companion study (fine-tune specialists from a shared base model, inspect task-vector compatibility, merge compatible checkpoints using TIES) but generalizes the workflow through an SE task registry covering nine classification tasks. Third, \textit{reproducibility by default}: DARE's random dropping is seedable, all merges are deterministic given the same seed, and bundled 200-row benchmark samples support smoke-test reproduction without requiring access to full benchmark datasets.
 
\noindent\textbf{SE Task Registry.} MergeSE includes a built-in registry of nine SE classification tasks: clone detection, vulnerability detection, defect prediction, code-smell detection, commit classification, code-review acceptability, comment-code consistency, exception-type prediction, and type inference. Each task entry specifies its canonical input shape (pair-input or single-input), default metric, expected CSV columns, and known benchmark formats. Users may also supply custom CSV datasets for tasks outside the registry. This registry allows MergeSE to auto-configure evaluation and detect input-format mismatches before inference begins.
 
\noindent\textbf{Five Operations.}
MergeSE exposes five operations through both the CLI and web interface. The \textit{tasks} operation lists registered SE task types and their metadata, giving users an overview of supported formats and evaluation defaults.
 
The \textit{inspect} operation diagnoses whether two or more checkpoints are suitable for merging. It checks base-model identity, tokenizer compatibility, and architectural compatibility, including model type, hidden size, and number of layers. When users provide a shared base checkpoint through the \texttt{--base} flag, MergeSE also computes pairwise task-vector cosine similarity and sign agreement, then reports a \textsc{compatible}, \textsc{risky}, or \textsc{incompatible} verdict. This diagnostic operationalizes the compatibility analysis from our companion study~\cite{roy2026unified}, where same-base checkpoints showed productive sign agreement while cross-base checkpoints approached near-random agreement. The verdict helps users avoid merge attempts that are structurally valid but unlikely to improve performance.
 
The \textit{merge} operation applies five merging algorithms, including TIES~\cite{yadav2023ties}, DARE-TIES~\cite{yu2024language}, Wudi~\cite{cheng2025wudi}, PCB~\cite{du2024parameter}, and simple task-vector averaging~\cite{ilharco2022editing}, to produce a unified checkpoint. If input checkpoints have incompatible classification heads, such as a binary clone-detection head and a multi-class commit-classification head, MergeSE automatically detects the mismatch and falls back to encoder-only merging. This preserves shared encoder knowledge while avoiding invalid arithmetic over task-specific output layers. The user can then attach or fine-tune an appropriate downstream head. This head-mismatch handling is important for SE workflows, where classification tasks often share an encoder architecture but differ in output dimensionality.
 
\begin{lstlisting}[language=bash,caption={Representative 
workflow: inspect, merge, and evaluate two CodeBERT clone 
detection specialists.},label={lst:workflow}]
$ mergese inspect --base microsoft/codebert-base \
    ./codebert_bcb ./codebert_clcdsa
  Verdict: COMPATIBLE
  Cosine sim: 0.077 | Sign agreement: 62.1%
 
$ mergese merge --base microsoft/codebert-base \
    -m ties ./codebert_bcb ./codebert_clcdsa \
    -o ./merged_codebert
  Merged in 4.9s | Trimmed: 22.3% | Conflicts: 15.5%
 
$ mergese evaluate ./merged_codebert \
    -t clone_detection -d ./bcb_test.csv
  F1: 0.801 | Precision: 0.747 | Recall: 0.864
\end{lstlisting}
 
The \textit{evaluate} operation loads a merged or specialist checkpoint and runs inference on a CSV benchmark. It reports F1, Precision, Recall, and Accuracy for binary tasks, and adds per-class breakdowns for multi-class tasks. MergeSE uses the task registry to determine whether the benchmark expects pair input, such as \texttt{code1, code2, label} for clone detection, or single input, such as \texttt{code, label} for vulnerability detection.
 
The \textit{export} operation packages a checkpoint for downstream deployment in three formats: HuggingFace, ONNX, and TorchScript. The HuggingFace export creates a Transformers-compatible directory for downstream loading. The ONNX export uses dynamic batch and sequence-length axes. The TorchScript export uses \texttt{torch.jit.trace}. Listing~\ref{lst:workflow} shows a complete inspect, merge, and evaluate workflow.
 
\noindent\textbf{Web Interface.}
In addition to the CLI, MergeSE provides a single-page web interface built with Flask, HTML, CSS, and JavaScript. The interface includes a Checkpoint Library with four input sources: HuggingFace Hub model IDs, uploaded checkpoint archives, admin-mounted server paths, and outputs from previous MergeSE jobs, enabling chained workflows where a merged checkpoint becomes input to a subsequent merge. Task cards for Inspect, Merge, Evaluate, and Export submit jobs to the backend, while Server-Sent Events stream logs to the browser in real time. The CLI (approximately 1,500 lines of Python) is the single source of truth; the web interface (approximately 3,200 lines) delegates all jobs via subprocess, ensuring identical behavior. Deployment uses Docker Compose or standard systemd configurations.
 
\noindent\textbf{Engineering Challenges.}
Building MergeSE exposed practical issues hidden in one-off research scripts. For example, \texttt{torch.quantile()} can fail on RoBERTa-style embedding matrices during TIES trimming; MergeSE uses \texttt{torch.kthvalue} as a replacement. Similarly, tokenizer version mismatches across training runs can cause loading failures; MergeSE implements a fallback chain to resolve compatible tokenizers. These decisions make MergeSE robust to heterogeneous checkpoint layouts common in SE research.
\section{Evaluation}
\label{sec:evaluation}

We evaluate whether MergeSE helps users diagnose, merge, evaluate, and deploy specialist SE models in a practical workflow. We use the same checkpoints and benchmarks as our companion study~\cite{roy2026unified} and reproduce the reported performance through MergeSE.

\noindent\textbf{RQ1: Can MergeSE diagnose checkpoint compatibility?}

\begin{table}[H]
\centering
\caption{MergeSE compatibility diagnostics \textsuperscript{$\dagger$}}
\label{tab:inspect}
\small
\begin{tabular}{lccc}
\toprule
\textbf{Checkpoint Pair} & \textbf{Verdict} &
\textbf{Cos.\ Sim.} & \textbf{Sign Agr.} \\
\midrule
CB-BCB $\leftrightarrow$ CB-CLCDSA & \textsc{compat.} &
+0.077 & 62.1\% \\
CB-BCB $\leftrightarrow$ UX-BCB & \textsc{incompat.} &
+0.102 & 53.2\% \\
\bottomrule
\end{tabular}
\\[2pt]
{\scriptsize $\dagger$ For cross-base pairs, geometry is reported only as a diagnostic because task vectors are not defined in a common parameter space. The verdict is therefore determined primarily by tokenizer and architecture incompatibility.}
\end{table}

Table~\ref{tab:inspect} shows that MergeSE identifies same-base CodeBERT checkpoints as merge-compatible and cross-base CodeBERT/UniXcoder checkpoints as incompatible. The same-base pair shows stronger sign agreement (62.1\%), while the cross-base pair is rejected because the checkpoints do not satisfy the tokenizer and architecture assumptions required for safe parameter merging. The inspect diagnostic helps users avoid merge-then-evaluate cycles that are structurally possible to attempt but unlikely to produce useful results.

\noindent\textbf{RQ2: Can MergeSE produce a useful merged model?}

\begin{table}[t]
\centering
\caption{MergeSE TIES output (mean $\pm$ std over 3 seeds) compared to individual specialists and multi-task training. Comb.\ is the arithmetic mean of BCB and CLCDSA F1.}
\label{tab:merge}
\small
\begin{tabular}{lcccc}
\toprule
\textbf{Model} & \textbf{BCB} & \textbf{CLCDSA} &
\textbf{Comb.} & \textbf{GPTCB} \\
\midrule
CB-BCB specialist & .894 & .464 & .679 & .450 \\
CB-CLCDSA specialist & .000 & .924 & .462 & .000 \\
UX Multi-task (all data) & \textbf{.933} & \textbf{.922} &
\textbf{.927} & .151 \\
\midrule
MergeSE TIES (CB) & .801{\tiny$\pm$.018} &
.746{\tiny$\pm$.012} & .774{\tiny$\pm$.014} &
\textbf{.609}{\tiny$\pm$.049} \\
MergeSE TIES (UX) & .890{\tiny$\pm$.010} &
.839{\tiny$\pm$.026} & .865{\tiny$\pm$.012} &
.450{\tiny$\pm$.042} \\
\bottomrule
\end{tabular}
\end{table}

Table~\ref{tab:merge} demonstrates the practical outcome of the MergeSE workflow on clone detection checkpoints from two model families. Individual specialists achieve strong in-domain performance but fail catastrophically on the other domain. For example, CodeBERT-CLCDSA scores 0.000 F1 on BigCloneBench. MergeSE TIES recovers cross-domain performance for both families: the CodeBERT merge achieves 0.774 combined F1, a 14\% improvement over the best individual CodeBERT specialist, while the UniXcoder merge achieves 0.865 combined F1, or 93\% of multi-task performance without using training data during merging. On unseen GPT-generated clones, the CodeBERT merge achieves the highest score across all configurations (0.609 F1), while the UniXcoder multi-task baseline collapses to 0.151 F1. This result illustrates the OOD trade-off that motivates post-hoc merging.

Validation against the reference merge from our companion study confirms numerical consistency: 149 of 199 parameter tensors are bitwise identical, and the remaining tensors show a median maximum absolute difference of $7.8 \times 10^{-3}$, with no observable effect on the reported evaluation metrics.

\noindent\textbf{RQ3: Is the tool efficient and deployable?}

\begin{table}[t]
\centering
\caption{MergeSE operation timing on CPU.}
\label{tab:timing}
\small
\begin{tabular}{llrl}
\toprule
\textbf{Operation} & \textbf{Input} & \textbf{Time} &
\textbf{Output} \\
\midrule
Inspect & 2 $\times$ ${\sim}$480\,MB & 7.2\,s &
Compat.\ report \\
TIES merge & 2 $\times$ 124M params & 4.9\,s &
HF checkpoint \\
Evaluate & ckpt + 200-row CSV & 17\,s &
F1/P/R report \\
Export (ONNX) & 1 checkpoint & 7.9\,s &
476\,MB \texttt{.onnx} \\
\bottomrule
\end{tabular}
\end{table}

Table~\ref{tab:timing} reports CPU wall-clock time on an Intel Xeon Platinum 8356H @ 3.90\,GHz with default PyTorch thread parallelism. The complete inspect, merge, evaluate, and export pipeline requires approximately 37 seconds. No GPU, training data, or retraining is required; evaluation needs only a labeled CSV.

\section{Related Work}
\label{sec:related}

\noindent\textbf{Model Merging Research.}
Task arithmetic~\cite{ilharco2022editing} introduced task vectors as a way to edit model behavior through parameter-space operations. TIES~\cite{yadav2023ties} improves task-vector merging by resolving sign conflicts through trimming and majority-sign election, while DARE~\cite{yu2024language} applies random dropping with rescaling to reduce redundant parameter updates. These techniques have been studied primarily in NLP and computer vision. To our knowledge, our companion study~\cite{roy2026unified} is the first systematic investigation of model merging for software engineering, showing that same-base task-vector merging can recover cross-domain clone detection performance without training data. MergeSE operationalizes these findings as a reusable tool for SE researchers.

\noindent\textbf{Clone Detection and OOD Generalization.}
Clone detection has long been studied across syntactic, semantic, and cross-language settings~\cite{roy2024exploratory}. Recent work shows that modern code-model detectors remain vulnerable to out-of-distribution generalization failures: in our companion study, individual specialists suffer 44--72\% F1 drops outside their training domain~\cite{roy2026unified}. Cross-language approaches such as CLCDSA~\cite{nafi2019clcdsa} and ZC3~\cite{li2023zc3} address specific generalization directions, but they require task-specific training objectives or retraining. Multi-task training can improve in-domain performance when all data are available, but our companion study shows that it can collapse on unseen clone types~\cite{roy2026unified}. MergeSE addresses the practical side of this problem by allowing researchers to combine existing specialist checkpoints post hoc, without training data or retraining.

\noindent\textbf{Merging Tools.}
mergekit~\cite{goddard2024arcee} is a general-purpose toolkit for large language model merging workflows, supporting methods such as linear interpolation, SLERP, TIES, and DARE. However, it is not designed around SE-specific encoder-model workflows, task registries, clone-pair evaluation, benchmark validation, or cross-task classification-head handling. Research codebases from the model merging literature also tend to provide experiment-specific scripts with hard-coded paths and benchmark-specific evaluation logic. MergeSE fills this tooling gap by providing an SE-aware merging workflow with compatibility diagnostics, an SE task registry, encoder-checkpoint support, cross-task head detection, built-in evaluation following SE benchmark conventions, and both CLI and web interfaces.

\section{Tool Availability and Roadmap}
\label{sec:roadmap}
MergeSE is publicly available at \textbf{\url{https://mergese.usask.ca}}, hosted on a server equipped with an NVIDIA L40 GPU to provide browser-based access without requiring local installation. Beyond model merging, MergeSE functions as a standalone evaluation workflow for software engineering (SE) classification tasks. Researchers can independently utilize the \textit{inspect}, \textit{evaluate}, and \textit{export} operations to diagnose checkpoint properties, execute standardized evaluations on SE benchmarks, and package models for deployment, all without writing dedicated evaluation code. This design establishes a reusable scientific workflow for SE model experimentation, standardizing routine operations such as benchmark evaluation, checkpoint comparison, and format conversion.

The current release includes built-in clone detection benchmarks, and users can evaluate custom SE classification tasks by supplying CSV datasets. Future releases will expand built-in templates to include vulnerability detection, defect prediction, and code-smell detection. Furthermore, our hosted infrastructure enables upcoming support for architecture-level merging via greedy layer stitching~\cite{roy2026unified}, which requires repeated validation-set inference during layer selection. We actively encourage community involvement through our GitHub repository at \textit{\url{https://github.com/srlabUsask/MergeSE}}, welcoming issue reports, benchmark contributions, and pull requests. To facilitate open-source contributions, MergeSE's modular CLI isolates each merging algorithm into a self-contained function operating directly on model state dictionaries.


\section{Conclusion}
\label{sec:conclusion}

This paper presented MergeSE, an open-source CLI and web tool 
that makes post-hoc model merging practical for software 
engineering. MergeSE turns what has historically been an 
experiment-specific script into a reusable workflow: from 
diagnosing whether two checkpoints are safe to merge, through 
the merge itself, to benchmark validation and deployment 
export. Our evaluation on clone detection checkpoints confirms 
that the tool produces merged models consistent with reference 
implementations and recovers cross-domain performance that no 
individual specialist achieves alone.

More broadly, we see MergeSE as addressing an emerging 
infrastructure gap in SE research. As pre-trained code models 
become the default starting point for tasks ranging from clone 
detection to vulnerability analysis, the number of fine-tuned 
specialist checkpoints will continue to grow. Without tooling 
for combining, evaluating, and deploying these checkpoints, 
each new specialist becomes another isolated artifact. MergeSE 
provides one path toward treating model composition as a 
routine part of the SE research workflow rather than a 
one-off experiment. We hope that by open-sourcing the tool 
and hosting it publicly, the SE community will contribute new 
merging algorithms, additional benchmarks, and extensions to 
new model families as they emerge.

\section*{Acknowledgements}
This research is supported in part by the Natural Sciences and Engineering Research Council of Canada (NSERC) Discovery Grants program, the Canada Foundation for Innovation's John R. Evans Leaders Fund (CFI-JELF), and by the industry-stream NSERC CREATE in Software Analytics Research (SOAR).


\bibliographystyle{ACM-Reference-Format}
\bibliography{bibliography_clean.bib}

\end{document}